# Impure codes exceeding the pure bounds for quantum local recovery[*]

Carlos Galindo[1][*][†], Fernando Hernando[1][*], Helena Martín-Cruz[2][*] and Ryutaroh Matsumoto[3][*]

[1][*]Instituto Universitario de Matemáticas y Aplicaciones de Castellón and Departamento de Matemáticas, Universitat Jaume I, Campus de Riu Sec., Castelló de la Plana, 12071, Castelló, Spain.
[2][*]Departamento de Matemáticas, Universidad de Jaén, Campus Las Lagunillas, 23071, Jaén, Spain.
[3][*]Department of Information and Communications Engineering, Institute of Science Tokyo, Ookayama 2-12-1, Meguro, 152-8550, Tokyo, Japan.

[*]Corresponding author(s). E-mail(s): galindo@uji.es; carrillf@uji.es; hmartin@ujaen.es; ryutaroh@ict.e.titech.ac.jp;
[†]ORCID: 0000-0002-3908-4462 (Galindo), 0000-0002-9758-2152 (Hernando), 0000-0002-6379-6902 (Martín-Cruz), 0000-0002-5085-8879 (Matsumoto)

### Abstract

Literature provides several bounds for quantum local recovery, which essentially consider the number of message qudits, the distance, the length, and the locality of the involved codes. We give a family of $J$-affine variety codes that result in impure CSS codes. These quantum codes exceed several of the above mentioned bounds that apply to pure quantum locally recoverable codes. We also discuss a connection between bounds on quantum local recovery and on weight-constrained stabilizer codes.

## 1 Introduction

An *erasure* in quantum and classical error correction means an error whose position in a codeword is known [2, 27, 37]. It is known that a quantum error-correcting code can correct twice as many erasures as errors. In light of this, recent papers [29, 44] take advantage of erasures in quantum fault-tolerant computation, as some physical devices allow identification of qubits with erasures in a codeword without destruction of encoded quantum information or stabilizer measurements [29, 44].

---

[*]Part of the results in this paper was submitted to WAIFI 2026 Santander as a short presentation without proceedings publication.

Stabilizer codes [1, 5, 6, 25, 30] are a class of quantum error-correcting codes allowing efficient implementation of encoders and decoders. Erasure correction involves measurements. Measurements are costly on some physical devices and measurement-free fault-tolerant computation has been actively investigated recently [28, 36, 41]. Reducing the number of measurements in quantum error correction has also been investigated [47]. In particular, measurements cause disturbance of measured qubits on some devices [28, 36, 41, 47]. In order to address this issue, in [42, 43] the authors studied lower and upper bounds on the number $k$ of message qubits and the distance $d$ realizable by a stabilizer group acting on $n$ qubits, whose generator weights are constrained by some bound, so that one can use stabilizers with small weights for quantum error correction.

As a somewhat related topic, classical locally recoverable codes with locality $r$ [24] are classical linear codes correcting a single erasure by referring to at most $r$ other codeword symbols. As an extension of [24], Prakash et al. [38] considered the correction of $\delta - 1$ ($\delta \geq 2$) erasures by referring to at most $r$ other codeword symbols, and proposed the concept of locally recoverable code with locality $(r, \delta)$. As quantum counterparts, quantum locally recoverable codes with locality $r$ were proposed by Golowich and Guruswami [22, 23] and those with locality $(r, \delta)$ were proposed by Galindo et al. [17]. Quantum local recovery can be regarded as another approach to reduce the number of measured qubits/qudits in quantum error or erasure correction.

The Singleton bound is central to classical coding theory and is closely linked to the MDS conjecture. Correspondingly, Singleton-like bounds have been established for quantum and entanglement-assisted quantum codes [3, 26, 30]. Establishing Singleton-like bounds (or other bounds) for quantum locally recoverable codes is a significant challenge. The Singleton-like bound for correction of a single erasure by general quantum locally recoverable error-correcting codes (2) is essentially an upper bound on $k$ in terms of $d$, $n$ and $r$, and was proposed in [22]. That for a single erasure corrected by CSS codes (3) was proposed in [34], and that for multiple erasures corrected by pure stabilizer codes (4) was proposed in [17]. Other bounds were also proposed and studied in [32, 33]. Li et al. [33] focused on pure quantum codes and obtained tighter bounds; see (5), (6) and (7). As noted, several bounds exist for single erasures, but only (4) is available for multiple erasures. Researchers use Singleton-like bounds to assess the optimality of quantum codes; indeed (2) was used in [4, 33, 34, 39, 45], (3) for the codes in [45], and (4) for the codes in [9, 17, 18, 46]. Also, any upper bound on quantum local recovery always implies an upper bound for weight-constrained stabilizers, as discussed in Remark 1.

In this paper, we provide a specific family of impure quantum locally recoverable codes exceeding the bounds (4), (5) and (6) for pure quantum codes.

The main ideas behind our code construction are as follows: Our family of linear codes $C$ resides in the intersection of $J$-affine variety codes [19], weighted Reed-Muller codes [21], and decreasing monomial-Cartesian codes [8]. We use $J$-affine variety codes [19] with empty $J$ to ensure that $C$ contains its dual $C^\perp$. We use decreasing monomial-Cartesian codes (or equivalently weighted Reed-Muller codes) to determine the true minimum Hamming distance of $C$. The locality of $C$ is determined by [16] as a monomial-Cartesian code. Finally the Feng-Rao bound [13] on the coset distance [11, 20] of the monomial-Cartesian codes [15] provides a lower bound on the minimum Hamming weight of $C \setminus C^\perp$ and ensures the impurity of the Calderbank-Shor-Steane (CSS) code [7, 40] constructed from $C$.

This paper is organized as follows: Section 2 reviews classical and quantum local recovery. The connection between quantum local recovery and weight-constrained stabilizers is also discussed there. In Section 3, we provide a family of quantum locally recoverable codes exceeding the bounds (4), (5) and (6) for pure quantum codes: here we use $J$-affine variety codes [19] with $J = \emptyset$. In Section 4 concluding remarks are given.

# 2 Preliminaries

## 2.1 Classical locally recoverable codes, general quantum locally recoverable codes and their Singleton-like bounds

An $[n, k, d]_q$ linear code is a $k$-dimensional linear space $C \subset \mathbf{F}_q^n$ with minimum Hamming distance $d$, where $q$ is a prime power and $\mathbf{F}_q$ is the finite field with $q$ elements. An $[[n, k, d]]_q$ quantum error-correcting code (QECC) is a $q^k$-dimensional complex subspace of $\mathcal{H}_q^{\otimes n}$ with distance $d$, where $\mathcal{H}_q$ is the $q$-dimensional complex linear space corresponding to the state space of unit quantum information, called a *qudit*. A QECC, $Q$, has distance $d$ if Pauli errors $E$ of weight less than $d$ are either detectable or $E|\varphi\rangle$ is a scalar multiple of $|\varphi\rangle$ for each $|\varphi\rangle \in Q$ but there exists a Pauli error $E'$ of weight $d$ that is undetectable and $E'$ changes some codeword in $Q$ [1, 6, 25, 30, 31]. A QECC $Q$ with distance $d$ is said to be impure if there exists some Pauli error $E$ of weight less than $d$ such that $E|\varphi\rangle$ is a scalar multiple of $|\varphi\rangle$ for each $|\varphi\rangle \in Q$, which implies that such an error $E$ is inevitably undetectable. If $Q$ is not impure, it is said to be pure.

As mentioned in Section 1, an erasure is an error whose position is known to a decoder [2, 27, 37]. A classical or quantum error-correcting code of length $n$ is said to have locality $(r, \delta)$ (or that it is $(r, \delta)$-locally recoverable) if for any index $i \in \{1, \ldots, n\}$ there exists a subset $i \in J \subseteq \{1, \ldots, n\}$ of size at most $r + \delta - 1$ such that any $\delta - 1$ erasures at $J$ can be corrected by a procedure only manipulating symbols at $J$. Formal definitions can be found in [17, 38].

An $[n, k, d]_q$ classical linear code with locality $(r, \delta)$ always satisfies the following inequality provided in [38]:

$$k + d + \left(\left\lceil \frac{k}{r} \right\rceil - 1\right)(\delta - 1) \leq n + 1. \tag{1}$$

An $[[n, k > 0, d]]_q$ QECC with locality $(r, \delta = 2)$ satisfies

$$k \leq n - 2(d-1) - \left\lfloor \frac{n - (d-1)}{r+1} \right\rfloor - \left\lfloor \frac{n - 2(d-1) - \left\lfloor \frac{n-(d-1)}{r+1} \right\rfloor}{r+1} \right\rfloor, \tag{2}$$

see [22, Theorem 35]. Note that (2) did not appear in the published proceedings paper [23].

*Remark 1* CSS codes will be reviewed in Section 2.2. Other stabilizer codes are discussed only in this remark, therefore we do not review them; the readers are referred to [1, 5, 6, 25, 30]. Wang et al. [42], Wei et al. [43] considered quantum error (or erasure) correction by stabilizers $S$ whose weight, denoted by $w(S)$ below, is constrained by some upper bound $w$. For an $n$-fold tensor product $M$ of Pauli matrices acting on $\mathcal{H}_q^{\otimes n}$, by the weight of $M$, $w(M)$, we denote the number of non-identity components in $M$. If $G$ is a finite set of tensor products $M$ as above, we set

$$w(G) := \max\{w(M) : M \in G\}, \text{ and}$$
$$w(S) := \min\{w(G) : G \text{ generates } S \text{ as a group of matrices}\}.$$

In this remark, assume $w(M) \neq 1$ for all $M \in S$. If there exists an $[[n, k, d \geq 2]]_q$ stabilizer code $Q$ defined by $S$ generated by $\{G_1, \ldots, G_\ell\}$ whose weights are not larger than $w$, then it has locality $2(w - 1)$ for correcting a single erasure. To show this, suppose that a Pauli erasure $E$ happens at index $i$ of a codeword. By assumption, $E \notin S$ and $E$ changes codewords in $Q$. Since the distance of $Q$ is larger than 1, there exist two generators $G_j$ and $G_{j'}$ whose measurements identify $E$. The matrices $G_j$ and $G_{j'}$ have non-identity matrices at the $i$-th index, and each of $G_j$ and $G_{j'}$ has at most $w - 1$ non-identity matrices at indices other than $i$. Thus, the total number of measured qudits other than the $i$-th one is at most $2(w - 1)$. This connection between quantum local recovery and weight-constrained stabilizers sheds new light on the importance of quantum local recovery. In addition, as far as the authors are aware, this connection has not been stated elsewhere. In Remark 3, we will show that the converse of this remark does not hold.

## 2.2 Quantum locally recoverable codes from the CSS construction and their Singleton-like bounds

Let $C$ be an $\mathbf{F}_q$-linear code of length $n$, and $C^\perp$ its dual code with respect to the standard Euclidean inner product [37]. If $C \supseteq C^\perp$, then one can construct an $[[n, 2\dim C - n, d_H(C \setminus C^\perp)]]_q$ quantum error-correcting code $Q(C)$ [1, 7, 30, 40], where $d_H(\cdot)$ denotes the minimum Hamming weight of nonzero vectors in $\cdot$, and $Q(C)$ is called the Calderbank-Shor-Steane (CSS) code, which is an important subclass of stabilizer codes [1, 5, 6, 25, 30]. A CSS code $Q(C)$ is pure if and only if $d_H(C) = d_H(C \setminus C^\perp)$ and impure otherwise.

The bound (2) holds for any QECC with a unitary encoding map. Luo et al. [34] gave another bound

$$2d \leq n - k - 2\left\lceil \frac{k}{r} \right\rceil + 4 \qquad (3)$$

for CSS codes. In contrast to (4) below, the bound (3) also applies to impure CSS codes. The bound (2) is asymptotically tighter than (3) [34, Eqs. (11) and (12)]. For pure CSS codes $Q$, if the parameters of $Q$ attain (3) with equality then they also attain (2) with equality [34, Theorem 9].

In [17], it was shown that the CSS code $Q(C)$ with parameters $[[n, k = (2\dim C - n), d_H(C \setminus C^\perp)]]_q$ has locality $(r, \delta)$ if $C$ has locality $(r, \delta)$. With the additional assumption $d = d_H(C) = d_H(C \setminus C^\perp)$, the CSS code $Q(C)$ with parameters

$$[[n, k = (2\dim C - n), d_H(C \setminus C^\perp)]]_q$$

has locality $(r, \delta)$ *only if* $C$ has locality $(r, \delta)$, which, by (1), implies

$$\frac{n+k}{2} + d + \left(\left\lceil \frac{n+k}{2r} \right\rceil - 1\right)(\delta - 1) \leq n + 1. \qquad (4)$$

Note that this is the only known bound for multiple erasures.

In Remark 1, we have shown that the existence of a quantum stabilizer code whose weight is constrained by a value $w$ implies that of a quantum locally recoverable code of locality $2(w - 1)$, correcting a single erasure. In our forthcoming Remark 3, we will see that the converse is not true. The following result will help.

**Lemma 2** *For every positive integer $m$, there exists a binary linear $[4m, 3m - 1]_2$-code $\mathcal{C}$ with locality 3 such that every parity check matrix $H$ for $\mathcal{C}$ always contains a row vector of Hamming weight at least $2m$.*

*Proof* Let us introduce a binary linear code $\mathcal{C}$ as in the statement. The code $\mathcal{C}$ of length $n = 4m$ over $\mathbf{F}_2$ is given by explicitly constructing its Euclidean dual space, $\mathcal{C}^\perp$. We express vectors $\vec{v} \in \mathbf{F}_2^n$ as $\vec{v} = (v_1, \ldots, v_m)$, where $v_i \in \mathbf{F}_2^4$, $1 \leq i \leq m$. For each $i$ as above, define $\vec{v}_i = (v_{i1}, \ldots, v_{im}) \in \mathbf{F}_2^n$ such that $v_{ii} = (1, 1, 1, 1)$ and the remaining $v_{ij}$ are (0,0,0,0). Also consider the vector $\vec{w} \in \mathbf{F}_2^n$ such that $w_i = (1, 1, 0, 0)$ for all $i$. Then, by definition

$$\mathcal{C}^\perp = \mathrm{span}(\vec{v}_1, \ldots, \vec{v}_m, \vec{w}).$$

It is straightforward to prove that the generators of $\mathcal{C}^\perp$ are mutually orthogonal and individually self-orthogonal. Therefore, $\mathcal{C}^\perp$ is self-orthogonal and, thus, $\mathcal{C} = (\mathcal{C}^\perp)^\perp$ is dual containing.

Let $H$ be a parity check matrix for $\mathcal{C}$. Its rows are a basis of $\mathcal{C}^\perp$. Since $\dim \mathcal{C}^\perp = m + 1$, and the $\mathbf{F}_2$-linear space generated by $\{\vec{v}_1, \ldots, \vec{v}_m\}$, $\langle \vec{v}_1, \ldots, \vec{v}_m \rangle$ has dimension $m$, there exists a row in $H$ whose corresponding vector $\vec{u}$ satisfies $\vec{u} \in \mathcal{C}^\perp \setminus \langle \vec{v}_1, \ldots, \vec{v}_m \rangle$. Thus

$$\vec{u} = \vec{w} + \sum_{i=1}^m a_i \vec{v}_i, \ \ a_i \in \mathbf{F}_2.$$

From our choice of $\vec{v}_i$ and $\vec{w}$, it is easy to deduce that the Hamming weight of each component $u_i$ of $\vec{u}$, $1 \leq i \leq m$, is 2, where, as before, $\vec{u} = (u_1, \ldots, u_m)$ and, thus, the Hamming weight of $\vec{u}$ is $2m$.

To conclude, note that the Hamming weight of each $\vec{v}_i$ is exactly 4. Because every single coordinate in the code is covered by exactly one of these weight-4 dual codewords, any erased symbol can be recovered by taking the sum of the other 3 symbols in its block. Therefore, the code has a constant locality of $r = 3$. This concludes the proof. $\square$

*Remark 3* The converse of our statement in Remark 1 is not true, that is, there is no function $W(r)$ such that the existence of a quantum locally recoverable code with locality $r$ implies that of a quantum stabilizer code whose stabilizer weight is constrained by $W(r)$. As a counterexample, consider the binary codes $\mathcal{C}$ introduced in Lemma 2. For every positive integer $m$, the code $\mathcal{C}$ has locality 3 and every parity check matrix $H$ of $\mathcal{C}$ always contains a row vector of Hamming weight at least $2m$. By [17], the associated CSS code $Q(\mathcal{C})$ has locality 3, but its stabilizer always has a generator whose weight is at least $2m$, proving the nonexistence of $W(r)$. Similar examples can be constructed for the non-binary case.

## 2.3 Quantum codes from linear dual-containing codes with respect to the Hermitian inner product

This section considers $\mathbf{F}_{q^2}$-linear codes $D \subset \mathbf{F}_{q^2}^n$. For $\vec{x} = (x_1, \ldots, x_n)$ and $\vec{y} = (y_1, \ldots, y_n) \in \mathbf{F}_{q^2}^n$, the Hermitian inner product between $\vec{x}$ and $\vec{y}$ is $x_1^q y_1 + \cdots + x_n^q y_n \in \mathbf{F}_{q^2}$. The dual code $D^{\perp_h}$ is the set of vectors $\vec{x} \in \mathbf{F}_{q^2}^n$ that are orthogonal to every $\vec{y} \in D$ with respect to the above Hermitian inner product.

We have the following construction, similar to that of CSS codes. If $D \supseteq D^{\perp_h}$, then one can construct an $[[n, 2\dim_{\mathbf{F}_{q^2}} D - n, d_H(D \setminus D^{\perp_h})]]_q$ quantum error-correcting code $Q(D)$, where $\dim_{\mathbf{F}_{q^2}} D$ denotes the dimension of $D$ as an $\mathbf{F}_{q^2}$-linear space [1, 6, 30]. The quantum code $Q(D)$ is pure if and only if $d_H(D) = d_H(D \setminus D^{\perp_h})$ and impure otherwise. If $C^\perp \subseteq C \subset \mathbf{F}_q^n$ and $D$ is the $\mathbf{F}_{q^2}$-linear space spanned by $C$, then we have $D \supseteq D^{\perp_h}$, $\dim_{\mathbf{F}_q} C = \dim_{\mathbf{F}_{q^2}} D$ and $d = d_H(C \setminus C^\perp) = d_H(D \setminus D^{\perp_h})$. Therefore, every bound on QECCs constructed by the Hermitian inner product also applies to the CSS codes.

Li et al. [33] proposed the following three bounds on QECCs constructed by the Hermitian inner product. All of them assume the purity of a QECC.

$$n \geq \max_{0 \leq \ell \leq \lceil (n+k)/(2r) \rceil - 1} \left\{ (r+1)\ell + \sum_{t=0, 2 | t}^{n+k-2r\ell-2} \lceil d/q^t \rceil \right\}, \tag{5}$$

$$d \leq \min_{0 \leq \ell \leq \lceil (n+k)/(2r) \rceil - 1} \left\{ q^{n+k-2r\ell-2} (q^2 - 1)(n - (r+1)\ell)/(q^{n+k-2r\ell} - 1) \right\}, \tag{6}$$

$$k \leq n - 2 \max_{0 \leq \ell \leq \lfloor (n-1)/(r+1) \rfloor} \left\{ \ell + \log_{q^2} \left( \sum_{i=0}^{\lfloor (d-1)/2 \rfloor} \binom{n - \ell(r+1)}{i} (q^2 - 1)^i \right) \right\}, \tag{7}$$

where $\ell$ is an integer in the above optimization problems.

# 3 Impure quantum locally recoverable codes exceeding previous bounds

Fix positive integers $H \geq 3$ and $V \geq 3$, and a prime power $q \geq 3$ such that both $H-1$ and $V-1$ divide $q-1$. Let $h = (H-1)/2$ and $v = (V-1)/2$. Also fix a primitive element $\alpha \in \mathbf{F}_q$, which means $\alpha^{q-1} = 1$ but $\alpha^i \neq 1$ for $i = 1, \ldots, q-2$. Let $x_i = 0$ for $i = 0$ and $x_i = \alpha^{i\frac{q-1}{H-1}}$ for $i = 1, \ldots, H-1$. Similarly, let $y_j = 0$ for $j = 0$ and $y_j = \alpha^{j\frac{q-1}{V-1}}$ for $j = 1, \ldots, V-1$. Consider the points in $\mathbf{F}_q^2$, $Q_{i,j} = (x_i, y_j)$, which determine the $H \times V$ grid:

$$P_{H,V} = \{Q_{i,j} : i = 0, \ldots, H-1 \text{ and } j = 0, \ldots, V-1\} \subseteq \mathbf{F}_q^2.$$

It is clear that $P_{H,V}$ is the Cartesian product of $\{x_0, \ldots, x_{H-1}\}$ and $\{y_0, \ldots, y_{V-1}\}$.

Let $I_{H,V}$ be the set of polynomials $F(X,Y)$ in the bivariate polynomial ring $\mathbf{F}_q[X,Y]$ such that $F(x_i, y_j) = 0$ for all $(x_i, y_j) \in P_{H,V}$. Since $\{x_0, \ldots, x_{H-1}\}$ are the roots of $X^H - X$ and those of $Y^V - Y$ are $\{y_0, \ldots, y_{V-1}\}$, $I_{H,V}$ is the ideal of $\mathbf{F}_q[X,Y]$ generated by $X^H - X$ and $Y^V - Y$. Moreover, $\{X^H - X, Y^V - Y\}$ forms a Gröbner basis [10, Definition 5 in p. 78] of the ideal $I_{H,V}$ with respect to any monomial order [10, Definition 1 in p. 55].

For integers $(H-1) \bmod 2 \leq a < h$ and $(V-1) \bmod 2 \leq b < v$, let us define

$$\Delta_{H,V,a,b} := \{X^i Y^j : 0 \leq j < v - b \text{ and } 0 \leq i \leq H-1, \text{ or } v - b \leq j \leq v + b \text{ and } 0 \leq i \leq h + a\}. \tag{8}$$

Then, the cardinality $|\Delta_{H,V,a,b}|$ of the set $\Delta_{H,V,a,b}$ equals

$$\lceil v - b \rceil H + (1 + \lfloor h + a \rfloor)(2b + (V \bmod 2)). \tag{9}$$

For a monomial $X^i Y^j \in \Delta_{H,V,a,b}$, let $\text{ev}(X^i Y^j) \in \mathbf{F}_q^{HV}$ be the evaluation of $X^i Y^j$ at the points in $P_{H,V}$.

Let $C(\Delta_{H,V,a,b})$ be the $\mathbf{F}_q$-linear code of length $HV$ spanned by the vectors $\text{ev}(X^i Y^j)$ for $X^i Y^j \in \Delta_{H,V,a,b}$.

Since the remainder of any monomial $M(X,Y)$ in $\Delta_{H,V,a,b}$ on division by the Gröbner basis $\{X^H - X, Y^V - Y\}$ [10, Theorem 3 in p. 64] is $M(X,Y)$ itself, the ($\mathbf{F}_q$-linear) map ev defined on the $\mathbf{F}_q$-linear space generated by the monomials $M(X,Y)$ is injective; this holds by the theory of Gröbner bases [10, Proposition 4 in p. 254] or that of the affine variety codes [14]. Consequently, the linear code $C(\Delta_{H,V,a,b})$ has parameters $[HV, |\Delta_{H,V,a,b}|]_q$.

Let us study the CSS code $Q(C(\Delta_{H,V,a,b}))$. Define

$$\Delta_{H,V,a,b}^{\perp} := \{X^i Y^j : 0 \leq j < v - b \text{ and } 0 \leq i \leq H-1, \text{ or } v - b \leq j \leq v + b \text{ and } 0 \leq i < h - a\}. \tag{10}$$

We observe that $\Delta_{H,V,a,b}^{\perp} \subsetneq \Delta_{H,V,a,b}$ and

$$\Delta_{H,V,a,b} \setminus \Delta_{H,V,a,b}^{\perp} = \{X^i Y^j : v - b \leq j \leq v + b \text{ and } h - a \leq i \leq h + a\} \neq \emptyset. \tag{11}$$

Note that $\Delta_{H,V,a,b} \setminus \Delta_{H,V,a,b}^{\perp}$ resides at the *center* of $\{X^i Y^j : 0 \leq i \leq H-1, 0 \leq j \leq V-1\}$, as visualized in Figs. 2 and 3. Since our classical linear code $C(\Delta_{H,V,a,b})$ is a particular instance of $J$-affine variety codes with empty $J$ as introduced in [19], by [19, Proposition 2] and (11) we have

$$C(\Delta_{H,V,a,b})^{\perp} = C(\Delta_{H,V,a,b}^{\perp}) \subsetneq C(\Delta_{H,V,a,b}),$$

which enables us to construct an $[[HV, (2a + (H \bmod 2))(2b + (V \bmod 2))]]_q$ CSS quantum code $Q(C(\Delta_{H,V,a,b}))$.

Next, we give two examples for a better understanding of the previous paragraphs.

*Example 4* Let $H = 5$, $V = 3$, $a = b = 0$ and $q = 5$. We have $h = 2$ and $v = 1$. A primitive element of $\mathbf{F}_5 = \{0, 1, 2, 3, 4\}$ can be chosen as $\alpha = 2$. Then $(x_0, \ldots, x_4) = (0, 2, 4, 3, 1)$, $(y_0, \ldots, y_2) = (0, 4, 1)$, and $P_{H,V}$ is the Cartesian product of $\{0, 2, 4, 3, 1\}$ and $\{0, 4, 1\}$ as

$$\begin{array}{ccccc} & & \overbrace{X^2Y}^{\notin \Delta_{H,V,a,b}^\perp} & & \\ Y & XY & X^2Y & & \\ 1 & X & X^2 & X^3 & X^4 \end{array}$$

|         |          |          |          |          |          |
|---------|----------|----------|----------|----------|----------|
| $y_2 = 1$ | $(0,1)$  | $(2,1)$  | $(4,1)$  | $(3,1)$  | $(1,1)$  |
| $y_1 = 4$ | $(0,4)$  | $(2,4)$  | $(4,4)$  | $(3,4)$  | $(1,4)$  |
| $y_0 = 0$ | $(0,0)$  | $(2,0)$  | $(4,0)$  | $(3,0)$  | $(1,0)$  |
|         | $x_0 = 0$ | $x_1 = 2$ | $x_2 = 4$ | $x_3 = 3$ | $x_4 = 1$ |

**Fig. 1** Points in $P_{H,V}$ with $H = 5$, $V = 3$ and $q = 5$ in Example 4

**Fig. 2** Monomials in $\Delta_{H,V,a,b}$ with $H = 5$, $V = 3$ and $a = b = 0$ in Example 4

$$\begin{array}{cccccccc} & & & \overbrace{Y^4X^3}^{\notin \Delta_{H,V,a,b}^\perp} & \overbrace{Y^4X^4}^{\notin \Delta_{H,V,a,b}^\perp} & & & \\ Y^4 & Y^4X & Y^4X^2 & Y^4X^3 & Y^4X^4 & & & \\ & & & \overbrace{Y^3X^3}^{\notin \Delta_{H,V,a,b}^\perp} & \overbrace{Y^3X^4}^{\notin \Delta_{H,V,a,b}^\perp} & & & \\ Y^3 & Y^3X & Y^3X^2 & Y^3X^3 & Y^3X^4 & & & \\ Y^2 & Y^2X & Y^2X^2 & Y^2X^3 & Y^2X^4 & Y^2X^5 & Y^2X^6 & Y^2X^7 \\ Y & YX & YX^2 & YX^3 & YX^4 & YX^5 & YX^6 & YX^7 \\ 1 & X & X^2 & X^3 & X^4 & X^5 & X^6 & X^7 \end{array}$$

**Fig. 3** Monomials in $\Delta_{H,V,a,b}$ with $H = V = 8$ and $a = b = 1$ in Example 5

shown in Fig. 1. As shown in Fig. 2, we see $\Delta_{H,V,a,b} = \{1, X, X^2, X^3, X^4, Y, XY, X^2Y\}$, $\Delta_{H,V,a,b} \setminus \Delta_{H,V,a,b}^\perp = \{X^2Y\}$, and then $C(\Delta_{H,V,a,b})$ is a $[15, 8]_5$ linear code.

*Example 5* Let $H = 8$, $V = 8$, $a = b = 1$ and $q = 8$. We have $h = v = \frac{7}{2}$. Let $\alpha$ be a primitive element of $\mathbf{F}_8$, where $\alpha^3 + \alpha + 1 = 0$. Then, $\mathbf{F}_8 = \{0, \alpha^0, \alpha^1, \ldots, \alpha^6\}$. Now,

$$(x_0, \ldots, x_7) = (y_0, \ldots, y_7) = (0, \alpha^0, \alpha^1, \ldots, \alpha^6),$$

and $P_{H,V}$ is the Cartesian product of $\{0, \alpha^0, \alpha^1, \ldots, \alpha^6\}$ and $\{0, \alpha^0, \alpha^1, \ldots, \alpha^6\}$. As shown in Fig. 3, we see that

$$\Delta_{H,V,a,b} = \{Y^j, XY^j, X^2Y^j, X^3Y^j, X^4Y^j, X^5Y^j, X^6Y^j, X^7Y^j \ : \ 0 \leq j \leq 2\}$$
$$\cup \{Y^3, XY^3, X^2Y^3, X^3Y^3, Y^4, XY^4, X^2Y^4, X^3Y^4, X^4Y^3, X^4Y^4\},$$

$\Delta_{H,V,a,b} \setminus \Delta_{H,V,a,b}^\perp = \{X^3Y^3, X^3Y^4, X^4Y^3, X^4Y^4\}$, and $C(\Delta_{H,V,a,b})$ is a $[64, 34]_8$ linear code.

For a monomial $X^iY^j$, by $d(X^iY^j)$ we denote its *distance* $(H - i)(V - j)$ introduced in [16, Definition 3.5]. Our classical linear code $C(\Delta_{H,V,a,b})$ is a particular instance of weighted Reed-Muller codes [21] and also of decreasing monomial-Cartesian codes [8, Section III], and by [21, Proposition 2] or [8, Theorem 3.9(iii)], we get

$$\begin{aligned} d_H(C(\Delta_{H,V,a,b})) &= \min_{X^iY^j \in \Delta_{H,V,a,b}} d(X^iY^j) \\ &= \min\{d(X^{\lfloor h+a \rfloor}Y^{\lfloor v+b \rfloor}), d(X^{H-1}Y^{\lceil v-1-b \rceil})\} \\ &= \min\{(H - \lfloor h + a \rfloor)(V - \lfloor v + b \rfloor), V - \lceil v - 1 - b \rceil\}. \end{aligned}$$

By [16, Proposition 3.10], $C(\Delta_{H,V,a,b})$ has locality

$$(V - \lceil v - b \rceil, \lceil v - b + 1 \rceil).$$

Finally, by [15, Definition 15 and Theorem 16] –considering the lexicographical order $\succcurlyeq$ with $Y \succcurlyeq X$– and (11), the minimum Hamming weight $d_H(C(\Delta_{H,V,a,b}) \setminus C(\Delta_{H,V,a,b})^\perp)$ is greater than or equal to

$$
\begin{aligned}
&d_H(C(\Delta_{H,V,a,b}) \setminus C(\Delta_{H,V,a,b})^\perp) \\
&\geq \min\{d(X^i Y^j) : X^i Y^j \in \Delta_{H,V,a,b} \text{ and there exists } X^{i'} Y^{j'} \in \Delta_{H,V,a,b} \setminus \Delta_{H,V,a,b}^\perp \\
&\quad \text{such that } X^i Y^j \succcurlyeq X^{i'} Y^{j'}\} \quad (12) \\
&= \min\{(H-i)(V-j) : X^i Y^j \in \Delta_{H,V,a,b} \text{ and there exists } X^{i'} Y^{j'} \in \Delta_{H,V,a,b} \setminus \Delta_{H,V,a,b}^\perp \\
&\quad \text{such that } X^i Y^j \succcurlyeq X^{i'} Y^{j'}\} \quad (13) \\
&= (H - \lfloor h+a \rfloor)(V - \lfloor v+b \rfloor). \quad (14)
\end{aligned}
$$

Note that the above Equality (14) is true due to a suitable choice of the set $\Delta_{H,V,a,b}$ and the monomial order $Y \succcurlyeq X$ used in application of [15, Theorem 16] to $C(\Delta_{H,V,a,b}) \setminus C(\Delta_{H,V,a,b})^\perp$, while (12) and (13) hold independently of monomial orders. Also, by following the argument in [21, Proposition 2] or [8, Theorem 3.9(iii)], we consider the polynomial

$$F(X,Y) = \prod_{i=0}^{\lfloor h+a \rfloor - 1}(X - x_i) \prod_{j=0}^{\lfloor v+b \rfloor - 1}(Y - y_j),$$

which was denoted by $F(X_1, \ldots, X_m)$ in the proof of [21, Proposition 2] and $f_\alpha$ in [8, Theorem 3.9(iii)]. The Hamming weight of $\mathrm{ev}(F(X,Y))$ is exactly $(H - \lfloor h+a \rfloor)(V - \lfloor v+b \rfloor)$. Since $\mathrm{ev}(F(X,Y))$ belongs to $C(\Delta_{H,V,a,b}) \setminus C(\Delta_{H,V,a,b})^\perp$, it follows that $d_H(C(\Delta_{H,V,a,b}) \setminus C(\Delta_{H,V,a,b})^\perp)$ is exactly $(H - \lfloor h+a \rfloor)(V - \lfloor v+b \rfloor)$. The preceding observations allow us to state (and prove) the following result:

**Proposition 6** *Keep the notation as introduced at the beginning of this section. Then, the CSS code $Q(C(\Delta_{H,V,a,b}))$ is an*

$$[[HV, (2a + (H \bmod 2))(2b + (V \bmod 2)), (H - \lfloor h+a \rfloor)(V - \lfloor v+b \rfloor)]]_q$$

*quantum locally recoverable code with locality $(V - \lceil v-b \rceil, \lceil v-b+1 \rceil)$. This CSS code is impure if and only if*

$$(H - \lfloor h+a \rfloor)(V - \lfloor v+b \rfloor) > V - \lceil v-1-b \rceil. \quad (15)$$

□

*Remark 7* As stated before, the bound (2) is asymptotically tighter than (3), see [34]. For $q = 3$ and values $H = V = 3$ and $a = b = 0$, we obtain a $[[9,1,4]]_3$ impure quantum locally recoverable code with locality $(2,2)$. The bound (2) holds with equality, while (3) holds with strict inequality. Our $[[9,1,4]]_3$ impure code indicates that the bounds (2) and (3) can differ at a very short code length.

The following proposition gives a condition under which the impurity condition (15) always holds.

**Proposition 8** *Keep the above notation and assume that $V$ is an odd integer and $b = 0$. Then, the quantum code $Q(C(\Delta_{H,V,a,b}))$ is impure.*

*Proof* We wish to prove the inequality:

$$(H - \lfloor h + a \rfloor)(V - \lfloor v \rfloor) > V - \lceil v - 1 \rceil = v + 2,$$

where the last equality holds because $v$ is an integer.

For a start, we know that $a < h$, which implies $2a < H - 1$ and then $2a \leq H - 2$. Now, $h + a = \frac{H-1+2a}{2}$ and therefore

$$h + a \leq \frac{H - 1 + H - 2}{2} = H - \frac{3}{2}.$$

Thus, $H - \lfloor h + a \rfloor \geq 2$ and then,

$$(H - \lfloor h + a \rfloor)(V - \lfloor v \rfloor) \geq 2(v + 1) > v + 2,$$

which concludes the proof. □

The next theorem provides a family of impure CSS codes violating the bound (4).

**Theorem 9** *With the above notation, suppose that $V$ is an odd integer and $b = 0$. Then, the code $Q(C(\Delta_{H,V,a,b}))$ is an impure quantum $(r, \delta)$-locally recoverable code with parameters*

$$[[HV, 2a + (H \bmod 2), (H - \lfloor h + a \rfloor)(v + 1)]]_q$$

*and $(r, \delta)$-locality $(v + 1, v + 1)$. These parameters and locality violate (4) for $H \geq 3$ and $((H - 1) \bmod 2) \leq a < h$.*

*Proof* It suffices to show that, with the parameters and locality as in the statement, the following inequality holds:

$$\frac{n + k}{2} + d + \left(\left\lceil \frac{n + k}{2r} \right\rceil - 1\right)(\delta - 1) > n + 1. \qquad (16)$$

We have $r = \delta = \frac{V+1}{2}$. Clearly, $k = 2a + 1$ if $H$ is odd and $k = 2a$ otherwise. Since $a < h$, $2a \leq H - 2$ and we get $H - k \geq 2 > 0$.

Let us give an expression for $d$ depending on $H$, $k$ and $r$. It is independent of the parity of $H$. Indeed, if $H$ is odd, one has $h = \frac{H-1}{2}$ and $a = \frac{k-1}{2}$ and then

$$d = (H - (h + a))r = \left(\frac{H - k}{2} + 1\right)r.$$

Otherwise, $h = \frac{H-1}{2}$, $a = \frac{k}{2}$ and $\lfloor h + a \rfloor = \lfloor \frac{H+k-1}{2} \rfloor = \frac{H+k}{2} - 1$. Thus,

$$d = \left[H - \left(\frac{H + k}{2} - 1\right)\right]r = \left(\frac{H - k}{2} + 1\right)r.$$

Replacing $d$ by the above value and noting that $\delta = r$ and $\lceil x \rceil \geq x$, one concludes that the left-hand side of (16) is larger than or equal to

$$n + k + \frac{H - k}{2}r - \frac{n + k}{2r} + 1.$$

Therefore, we only need to show that $k+\frac{H-k}{2}r-\frac{n+k}{2r}$ is strictly positive. Clearing denominators and substituting $n = HV$ and $r = \frac{V+1}{2}$, we desire to show the following inequality

$$kV + (H-k)\frac{V^2+2V+1}{4} - HV > 0.$$

It holds if and only if

$$(H-k)(V-1)^2 > 0 \qquad (17)$$

and the proof is completed because we previously proved that $H - k > 0$. □

*Remark 10* Considering the family of codes treated in Theorem 9, the value $\frac{(H-k)(V-1)^2}{8r}$ is a lower bound for the difference between the left-hand side and the right-hand side in (4) –see (17)–. This shows that by enlarging $q$, the difference can become arbitrarily large. Therefore, the addition of a constant to the right-hand side in (4) does not make (4) applicable to impure quantum $(r,\delta)$-locally recoverable codes.

Our next result shows that (5) can also be violated by impure codes.

**Theorem 11** *Keep the above notation and assume $V = 3$ and $b = 0$. Then the code $Q(C(\Delta_{H,V,a,b}))$ is an impure quantum locally recoverable code with parameters*

$$[[3H, 2a + (H \bmod 2), 2(H - \lfloor h + a \rfloor)]]_q$$

*and $(r, \delta)$-locality $(2, 2)$. If $H = q$ is a prime power, then its parameters and locality $r = 2$ violate (5) for every odd integer $H \geq 3$ and $0 \leq a < h$.*

*Proof* Continuing to use the previous notation, we deduce from the statement that $q$ is an odd prime power because $V - 1 = 2$ divides $q - 1$. Moreover, $v = 1$, $r = 2$, $n = 3q$, $k = 2a + 1$ and $d = q + 1 - 2a$. Thus, to violate (5), we have to prove that

$$3q < \max_{0 \leq \ell \leq \lceil \frac{3q+2a+1}{4} \rceil - 1} \left\{ 3\ell + \sum_{\substack{t=0 \\ 2|t}}^{3q+2a-4\ell-1} \left\lceil \frac{d}{q^t} \right\rceil \right\}. \qquad (18)$$

The fact that $\ell \leq \frac{3q+2a-1}{4}$ implies that the upper limit of the sum in (18) is not negative.

Assume $t \geq 2$. By hypothesis $0 \leq a \leq h - 1 = \frac{q-3}{2}$ and $0 < d \leq q + 1 \leq 3q \leq q^2$. Then, $0 < d/q^t \leq d/q^2 < 1$ and, therefore, $\lceil d/q^t \rceil = 1$. We have proved that

$$3\ell + \sum_{\substack{t=0 \\ 2|t}}^{3q+2a-4\ell-1} \left\lceil \frac{d}{q^t} \right\rceil = 3\ell + (q+1-2a) + \frac{3q+2a-4\ell-1}{2} = \ell + \frac{5q+1}{2} - a,$$

which is strictly increasing with respect to $\ell$. This proves that

$$\left\lceil \frac{3q+2a+1}{4} \right\rceil - 1 + \frac{5q+1}{2} - a$$

is the right-hand side of (18).

As a consequence, to conclude the proof it suffices to show that

$$\frac{q+1+2a}{2} < \left\lceil \frac{3q+2a+1}{4} \right\rceil. \qquad (19)$$

To see it, note that
$$\frac{3q+2a+1}{4} = \frac{q+1+2a}{2} + \frac{q-2a-1}{4}$$
and since $a \leq (q-3)/2$, $q-2a-1 \geq 2$ and $\frac{q-2a-1}{4} \geq \frac{1}{2} > 0$ which, after noting that $\frac{q+1+2a}{2}$ is an integer, proves (19) and the theorem. □

Violation of the bound (6) is treated in our last theorem.

**Theorem 12** *With the above notation, let $V = 3$, $H = 2h+1 \geq 3$ be an odd integer, and $a = b = 0$. Then, the CSS code $Q(C(\Delta_{H,V,a,b}))$ is an impure quantum locally recoverable code with parameters $[[3(2h+1), 1, 2(h+1)]]_q$ and $(r,\delta)$-locality $(2,2)$. If $H = q$ is a prime power, then its parameters and locality, $r = 2$, violate (6).*

*Proof* For proving the theorem, we have to show that

$$d > \min_{0 \leq \ell \leq \lceil (n+k)/(2r) \rceil - 1} \left\{ \frac{q^{n+k-2r\ell-2}(q^2-1)(n-(r+1)\ell)}{q^{n+k-2r\ell}-1} \right\}, \tag{20}$$

where $n = 3q$, $k = 1$, $d = q+1$ and $r = 2$.

Consider the map $\phi : [0, \lceil \frac{3q+1}{4} \rceil - 1] \cap \mathbf{Z} \to \mathbf{Q}$ defined by

$$\phi(\ell) := \frac{3q^{3q-4\ell-1}(q^2-1)(q-\ell)}{q^{3q-4\ell+1}-1},$$

where $\mathbf{Z}$ and $\mathbf{Q}$ denote the sets of integers and rational numbers, respectively. We observe that $\phi$ is monotonically decreasing with $\ell$. Indeed, to show this, it suffices to prove that

$$\frac{\phi(\ell)}{\phi(\ell+1)} > 1 \quad \text{for} \quad 0 \leq \ell \leq \left\lceil \frac{3q+1}{4} \right\rceil - 2.$$

This inequality is equivalent to

$$\frac{q-\ell}{q-\ell-1} > \frac{q^{3q-4\ell+1}-1}{q^{3q-4\ell+1}-q^4},$$

which can be written as

$$1 + \frac{1}{q-\ell-1} > 1 + \frac{q^4-1}{q^{3q-4\ell+1}-q^4}.$$

Therefore we must prove

$$q^{3q-4\ell+1} - q^4 > (q^4-1)(q-\ell-1)$$
$$\Leftrightarrow q^{3q-4\ell+1} > q^5 - q^4\ell - q + \ell + 1. \tag{21}$$

For $\ell = 0$ we see (21) holds as $q \geq 2$. For $\ell > 0$, we have $\ell \leq \lceil \frac{3q+1}{4} \rceil - 2$, which means $3q - 4\ell + 1 > 4$, ensuring that (21) holds.

As a consequence of $\phi$ being monotonically decreasing, the right-hand side of (20) is equal to $\phi(\lceil \frac{3q+1}{4} \rceil - 1)$ and thus, to conclude the proof, we have to prove the following inequality:

$$d > \phi\left(\left\lceil \frac{3q+1}{4} \right\rceil - 1\right).$$

Because $q \geq 3$ is an odd prime power, $q$ must satisfy either $q \equiv 3 \pmod 4$ or $q \equiv 1 \pmod 4$.

Let us start with the case $q \equiv 3 \pmod 4$. Here,

$$\left\lceil \frac{3q+1}{4} \right\rceil - 1 = \frac{3q+3}{4} - 1 = \frac{3q-1}{4}$$

and, therefore,

$$\phi\left(\frac{3q-1}{4}\right) = \frac{3q^0(q^2-1)\left(q - \frac{3q-1}{4}\right)}{q^2-1} = 3\left(\frac{q+1}{4}\right) = \frac{3}{4}(q+1) < q+1 = d.$$

It remains to consider the case $q \equiv 1 \pmod 4$, which only is satisfied for prime powers larger than 3. Now,

$$\left\lceil \frac{3q+1}{4} \right\rceil - 1 = \frac{3q+1}{4} - 1 = \frac{3q-3}{4}.$$

Thus,

$$\phi\left(\frac{3q-3}{4}\right) = \frac{3q^3+9q^2}{4(q^2+1)}$$

and we must check if $q+1 > \frac{3q^3+9q^2}{4(q^2+1)}$, which is equivalent to $q^3 - 5q^2 + 4q + 4 > 0$. This inequality can be written as $q^2(q-5) + 4q + 4 > 0$, which is true for any prime power $q$ larger than 3. This concludes the proof. □

*Remark 13* Bounds (2), (3) and (7) hold for our family of impure codes for all odd integers $H \geq 3$ and $V \geq 3$, all $0 \leq a \leq h-1$ and $b = 0$.

Let us give an example showing how the above mentioned bounds are violated.

*Example 14* We continue to use the values $H = 5$, $V = 3$ and $a = b = 0$ from Example 4. The code $C(\Delta_{H,V,a,b})$ is a $[15, 8, 3]_5$ linear code with locality $(2, 2)$ and the distance of the corresponding CSS code $Q(C(\Delta_{H,V,a,b}))$ is $d_H(C(\Delta_{H,V,a,b}) \setminus C(\Delta_{H,V,a,b})^\perp)$, which is equal to 6. The difference between $d_H(C(\Delta_{H,V,a,b}) \setminus C(\Delta_{H,V,a,b})^\perp)$ and $d_H(C(\Delta_{H,V,a,b}))$ is 3, which is quite large compared with the code length $n = 15$.

Consider the polynomial $F(X, Y) = X^4 - 1 \in \mathbf{F}_5[X, Y]$. The components in the evaluation vector $\operatorname{ev}(F(X,Y))$ are nonzero exactly at $(0, 0)$, $(0, 1)$ and $(0, 4)$. Then, $\operatorname{ev}(F(X,Y))$ has Hamming weight 3 and belongs to $C(\Delta_{H,V,a,b})^\perp \subsetneq C(\Delta_{H,V,a,b})$. Since $\operatorname{ev}(F(X,Y)) \in C(\Delta_{H,V,a,b})$, we see that it is undetectable. Since $\operatorname{ev}(F(X,Y)) \in C(\Delta_{H,V,a,b})^\perp$, it keeps every quantum codeword in $Q(C(\Delta_{H,V,a,b}))$ unchanged. The existence of $\operatorname{ev}(F(X,Y)) \in C(\Delta_{H,V,a,b})^\perp$ makes the CSS code $Q(C(\Delta_{H,V,a,b}))$ impure.

With these parameters $n = 15$, $k = 1$, $d = 6$, $r = \delta = 2$, the pure Singleton-like bound (4) reduces to

$$17 \leq 16,$$

the pure Griesmer-like bound (5) reduces to

$$15 \geq 16,$$

and the pure Plotkin-like bound (6) reduces to

$$6 \leq \frac{75}{13},$$

which exemplify Theorems 9, 11 and 12.

Finally, with the above values for $H, V, a$ and $b$, $C(\Delta_{H,V,a,b})$ is a $[15, 8, 3]_5$ linear code with locality $(2, 2)$. The left-hand side of the classical Singleton-like bound (1) is 14 while the right-hand side is 16. This example shows that the quantum Singleton-like bound (4) can be exceeded by an impure quantum code constructed from a classical locally recoverable code that fails to attain the classical Singleton-like bound (1).

This section concludes with a new example and a remark.

*Example 15* Now, we use the values $H = V = 8$ and $a = b = 1$ from Example 5. The linear code $C(\Delta_{H,V,a,b})$ has parameters $[64, 34, 6]_8$ and $(r, \delta)$-locality $(5, 4)$. The distance of the corresponding CSS code $Q(C(\Delta_{H,V,a,b}))$ is 16. The number of message qudits is $34 \cdot 2 - 64 = 4$.

With these parameters $n = 64$, $k = 4$, $d = 16$, $r = 5$, $\delta = 4$, the pure Singleton-like bound (4) reduces to
$$68 \leq 65.$$
The pure bounds (5), (6) or (7) do not consider the multiple erasure case.

This example is not covered by Proposition 8, but the obtained CSS code is impure. Similarly, it is not covered by Theorem 9 but still exceeds the pure Singleton-like bound. Actually, one can determine and prove a necessary and sufficient condition in terms of $H$, $V$, $a$ and $b$ for the pure Singleton-like bound (4) to fail with our codes' family, but it is omitted because the condition is somewhat complicated.

*Remark 16* In our construction of the impure code $Q(C(\Delta_{H,V,a,b}))$, we evaluated what is called the coset distance $d_H(C(\Delta_{H,V,a,b}) \setminus C(\Delta_{H,V,a,b})^\perp)$ or the first relative generalized Hamming weight [35]. Our bounding technique originated from the Feng-Rao bound on primary codes [20]. The connection between the Feng-Rao bound [13] and the coset distance has been known since the last century [11]. The coset distance on various algebraic codes has been studied by the Feng-Rao bounding technique, for example, in [12]. By using this kind of bounding technique, one could expect to find other impure quantum locally recoverable codes exceeding the Singleton-like bounds.

## 4 Concluding Remarks

In this paper, we provided a family of $J$-affine variety codes [19] resulting in impure quantum locally recoverable codes by the CSS construction, violating several bounds for pure quantum locally recoverable codes (4), (5) and (6). One might wonder if one can make (4) applicable also to impure stabilizer codes by adding some constant to the right-hand side of (4). In Remark 10 we discussed the impossibility of such an approach.

There are several bounds applicable to impure QECCs [22, 23, 32, 34], all of which assume a single erasure $\delta = 2$. It is an important open problem to find upper bounds applicable to impure stabilizer codes and general QECCs for correcting multiple erasures ($\delta \geq 3$). When the CSS code $Q(C)$ with distance $d$ and locality $r_Q$ is constructed from a classical linear code $C$, $Q(C)$ becomes impure if there exists an undetectable nonzero error vector $\vec{e} \in C$ whose Hamming weight is strictly less than $d_H(C \setminus C^\perp)$. The existence of such an $\vec{e}$ makes $d$ different from the Hamming distance $d_H(C)$, as shown in Proposition 6 and Example 14. There might exist a situation where such an $\vec{e}$ also makes $r_Q$ different from the locality $r$ of the classical locally recoverable code $C$. Exploration of bounds for impure QECCs locally correcting multiple erasures seems nontrivial because we must consider differences between quantum and classical distances and localities.

**Acknowledgments.** This work was partially funded by MICIU/AEI/10.13039/-501100011033, by ERDF, UE (grant PID2022-138906NB-C22), and by the Japan Society for Promotion of Science under Grant No. 23K10980.

**Data availability.** No datasets were generated or analyzed during the current study.

**Declarations.** The authors have no competing interests to declare that are relevant to the content of this paper.

**Use of AI.** Initial proofs of Lemma 2, and Theorems 9, 11 and 12 were obtained with the help of AI (Google Gemini 3.1). They were completely rewritten by the authors, and no AI-generated texts or equations remain in this paper.